\RequirePackage{fix-cm}
\documentclass[twocolumn,english,sort&compress]{svjour3}          
\smartqed  
\usepackage{graphicx}

\usepackage[unicode=true,
 bookmarks=true,bookmarksnumbered=false,bookmarksopen=false,
 breaklinks=false,pdfborder={0 0 0},backref=false,colorlinks=false]
 {hyperref}
\usepackage[numbers]{natbib}
\usepackage{cite}

\usepackage{subfig}

\begin{document}

\title{Optimizing performance-per-watt on GPUs in High Performance Computing\thanks{
The authors acknowledge support from NSF grants PHYS-080357, AST-1106059,
and OIA-1120587. BB thanks the NVIDIA internship program for support.} 
}
\subtitle{Temperature, frequency and voltage effects}


\author{D.~C.~Price   \and
	M.~A.~Clark \and
        B.~R.~Barsdell \and
        R.~Babich \and
        L.~J.~Greenhill
}


\institute{D. ~C.~Price, B.~R.~Barsdell, L.~J.~Greenhill \at
             Harvard-Smithsonian Center for Astrophysics, MS 42, 60 Garden Street, Cambridge MA 01238 USA \\
              \email{dprice@cfa.harvard.edu}           
           \and
           M.~A.~Clark, R.~Babich \at
              NVIDIA, 2701 San Tomas Expy, Santa Clara, CA 95050 USA
}

\date{Received: date / Accepted: date}

\maketitle

\begin{abstract}
The magnitude of the real-time digital signal processing challenge
attached to large radio astronomical antenna arrays motivates use
of high performance computing (HPC) systems. The need for high power
efficiency  at remote observatory sites parallels
that in HPC broadly, where efficiency is a critical metric.
We investigate how the performance-per-watt of graphics processing
units (GPUs) is affected by temperature, core clock frequency and
voltage. Our results highlight how the underlying physical processes
that govern transistor operation affect power efficiency. In particular,
we show experimentally that GPU power consumption increases non-linearly
(quadratic) with both temperature and supply voltage, as predicted by physical
transistor models. We show lowering GPU supply voltage and increasing
clock frequency while maintaining a low die temperature increases
the power efficiency of an NVIDIA K20 GPU by up to 37-48\% over default
settings when running xGPU, a compute-bound code used in radio astronomy.
We discuss how automatic temperature-aware
and application-dependent voltage and frequency scaling (T-DVFS and
A-DVFS) may provide a mechanism to achieve better power efficiency
for a wider range of compute codes running on GPUs.
\end{abstract}

\section{Introduction}

Power efficiency is a crucial design factor within HPC. Power consumption is often
a limiting factor for HPC systems, with current generation 
machines already requiring power budgets $>$1~MW to operate\footnote{\href{http://www.green500.org/lists/green201411}{http://www.green500.org/lists/green201411}%
}. 

In order to build Exascale systems ($>10^{18}$ floating point operations per second, i.e. FLOPS), increasing
the achieved performance-per-watt of HPC hardware is of paramount importance 
for several reasons. Firstly and foremostly, the more power consumed, the more the system costs to
operate. Secondly, generation and distribution of power is non-trivial
on megawatt scales. In addition, waste heat poses an engineering challenge: 
it must be removed to avoid compute nodes overheating and failing. HPC cooling systems
often require significant amounts of power themselves. 
Decreasing compute power consumption in turn decreases infrastructure
power consumption and as such is the most promising way to increase
overall power efficiency.

Machines based upon graphic processing units (GPUs) dominate the Green 500
list, with 9 of the top 10 machines featuring GPUs. 
Indeed, two of the top 10 most powerful computers on
the June 2014 Top 500 list%
\footnote{\href{http://www.top500.org/lists/2014/11/}{http://www.top500.org/lists/2014/11/}%
}, Titan (2nd) and Piz Daint (6th), utilize Kepler GPUs. As such,
the question of how best to increase GPU power efficiency is pressing. 

In this paper, we investigate how performance-per-watt can be optimized
for an NVIDIA K20 GPU. We approach the problem by considering the
physical processes that govern transistor performance; in particular,
how temperature, supply voltage, and clock frequency affect power
efficiency.

\subsection{Power efficiency and GPUs in radio astronomy }

The proposed all-sky imaging element of the Long Wavelength Array
(LWA) \citep{Ellingson:2013ko}, with $\sim0.1$km$^{2}$ collecting
area and the proposed Square Kilometre Array (SKA) telescope%
\footnote{\href{http://www.skatelescope.org}{http://www.skatelescope.org}%
} will demand computation at peta- and exa-scale processing respectively
(e.g. \citep{Broekema:2012hw}). 
Due to the number of computations required, power efficiency is of particular concern for 
these and other next-generation radio telescopes. Constrained operational budgets further dictate
that strict power usage targets must be met.
In the past, custom hardware has been built to perform the
required digital signal processing tasks. For example, the VLA WIDAR
and ALMA correlators \citep{Perley:2009fp,Wootten:2009hw} are very
capable and power efficient signal processing systems, but they lack
the flexibility afforded by architectures where processing is performed
on general purpose computing platforms; they also took over a decade
to design and implement. It has been shown that GPU-based signal processing
systems for radio astronomy can be designed and deployed in a fraction
of this time, see for example \citep{Kocz:2014jr}.

GPUs are well-suited to many of the signal processing tasks required
in radio astronomy.
If power efficiency challenges can be met, then a GPU-based HPC system
would be an attractive solution for SKA signal processing. A GPU-based implementation of the SKA1-Low
central signal processor (a subsystem of the full SKA) is calculated
to require $\sim$335 kW, based on current NVIDIA Kepler GPU architecture \citep{Magro:2014ul}.
This assumes a GPU power efficiency of 12 GFLOPS/W;
we report 18.3 GFLOPS/W for the \texttt{xGPU} cross-correlation code,
after temperature-aware tuning of GPU supply voltage and frequency.
Whether or not GPUs are considered for SKA1-Low signal processing
hardware will depend upon demonstrating that GPUs can achieve acceptable
power efficiency within the next few years.

\subsection{Power leakage }

The physical processes that
underlie power usage are common across all architectures. These processes
can be broadly broken into two categories: static and dynamic. That
is, total power usage $P_{\mathrm{sys}}$ is given by
\begin{equation}
P_{\mathrm{sys}}=P_{\mathrm{static}}+P_{\mathrm{dynamic}}.\label{eq:pow-tot}
\end{equation}
The dynamic power is the power consumed in switching logic states,
given for a single logic component by 
\[
P_{\mathrm{dynamic}}=CV_{\mathrm{dd}}^{2}f_{\mathrm{clock}},
\]
 where $C$ is the load capacitance, $V_{\mathrm{dd}}$ is the voltage
swing and $f_{\mathrm{clock}}$ is the switching frequency. For a
chip with many logic components, the dynamic power is the sum of the
contributions of all $N_{\mathrm{c}}$ components:
\begin{equation}
P_{\mathrm{dynamic}}=\sum_{\mathrm{n}=1}^{N_{\mathrm{c}}}C_{\mathrm{n}}V_{\mathrm{n}}^{2}f_{\mathrm{n}},\label{eq:pow-dyn}
\end{equation}
 which for devices with a single clock domain (i.e. switching frequency),
voltage swing $V_{\mathrm{dd}}$, and identical logic components simplifies
to $N_{\mathrm{c}}CV_{\mathrm{dd}}^{2}f_{\mathrm{clock}}$. 

Static power, also known as \emph{leakage power}, is consumed regardless
of transistor switching and is due to current leakage;
more detailed discussion of these mechanisms can be found in \citep{Liao:2005ft,Liu:2007ga}. 
For sub-micrometer processes (i.e. most current-generation compute
architectures), subthreshold leakage is the dominant mechanism.

It is informative to consider an analytical expression for subthreshold
leakage. As shown in \citep{Liu:2007ga}, $I_{\mathrm{sub}}$ of a
MOS device can be expressed as
\begin{equation}
I_{\mathrm{sub}}=A_{\mathrm{s}}\frac{W}{L}\left(\frac{kT}{q}\right)^{2}e^{\frac{q(V_{\mathrm{gs}}-V_{\mathrm{thr}})}{nkT},}\label{eq:sub-thr}
\end{equation}
where $A_{{\rm s}}$ is a technology-dependent constant, 
$W$ and $L$ are device's effective channel width and length,
 $V_{\mathrm{gs}}$ is gate-to-source voltage,  
and $n$ is
the transistor's subthreshold swing coefficient. The quantity $kT/q$
is the thermal voltage, where $k$ is Boltzmann's constant, $q$ is
electron charge, and $T$ is temperature. The threshold
voltage $V_{{\rm thr}}$ is also a (non-linear) function of temperature,
decreasing with increasing temperature. 

Eq.~\ref{eq:sub-thr} predicts that subthreshold leakage current
exhibits a non-linear temperature dependence, proportional to $I_{{\rm sub}}\propto T^{2}e^{-b/T}$,
where $b>0$. Here, the exponent is necessarily
negative, as $(V_{{\rm GS}}-V_{\mathrm{thr}})<0$ (by definition of
subthreshold), $q/k\approx11605$ K/V, and $n\ge1$;
it follows that Eq.~\ref{eq:sub-thr} monotonically increases
with temperature. This implies that power efficiency of a transistor
increases with decreasing temperature. One therefore expects to see performance-per-watt of GPUs
improve as die temperature is lowered.

\subsection{Maximizing power efficiency\label{sub:Maximizing-power-efficiency}}

Maximizing power efficiency ($\eta_{\mathrm{pow}}$), requires simultaneous
optimization of power consumption and computational performance. In tension with
Eq.~1-3, compute performance ($N_{\mathrm{OPS}}$), 
increases linearly with clock frequency. That is, maximum power efficiency is given by 
\begin{equation}
\eta_{\mathrm{pow}}=\frac{N_{\mathrm{OPS}}}{P_{\mathrm{total}}}=\frac{N_{\mathrm{OPS}}}{P_{\mathrm{dynamic}}+P_{\mathrm{static}}}.
\end{equation}
For a simple chip with full utilization of $N_{\mathrm{c}}$ identical
compute components, each performing one operation per clock cycle,
$ N_{\mathrm{OPS}}=N_{\mathrm{c}}f_{\mathrm{clock}}$, where $f_{\mathrm{clock}}$
is the clock frequency,
\begin{equation}
\eta_{\mathrm{pow}}=\frac{N_{\mathrm{c}}f_{\mathrm{clock}}}{N_{\mathrm{c}}CV_{\mathrm{dd}}^{2}f_{\mathrm{clock}}+P_{\mathrm{static}}}.\label{eq:pow-eff}
\end{equation}
Eq.~\ref{eq:pow-eff} shows that power efficiency is increased
when voltage is decreased. Due to the $P_{\mathrm{static}}$ term
in the denominator, efficiency also increases with clock frequency.

For a complex chip such as a GPU this formalism is a simplification.
Another consideration is that frequency and
voltage are generally scaled together, not separately. This is primarily
as the speed at which a digital circuit can switch states from low
to high --- the gate delay time $t_{\mathrm{delay}}$ --- is 
\begin{equation}
t_{\mathrm{delay}}\propto\frac{V_{\mathrm{dd}}T^{\mu}}{(V_{\mathrm{dd}}-V_{\mathrm{thr}})^{\xi}},\label{eq:tdel-switch}
\end{equation}
where $\xi$ and $\mu$ are technology-dependent constants \citep{Liao:2005ft}.
The temperature dependence of Eq.~\ref{eq:tdel-switch}
arises as temperature affects carrier mobility and threshold voltage.
At higher frequencies, there is more dynamic power usage, so die temperature
will in turn increase, forcing higher $V_{\mathrm{dd}}$ to maintain
suitable $t_{\mathrm{delay}}$ (which in turn increases power usage
and die temperature).

Nonetheless, the default clock-voltage combination has been shown
to be conservative on some GPUs (see \citep{Mei:2013dk, leng2014}). 


\subsection{GPU power measurement and modeling}

The simplified power efficiency formula presented in Eq.~\ref{eq:pow-eff}
is not immediately applicable to GPUs, which feature hierarchical
memory, different clock domains, multiple instructions, and dynamic
control of voltage and clock frequency (DVFS). As such, there have
been many analyses at higher abstraction levels that quantify the
power characteristics of GPU hardware and provide models that predict
power usage \citep{Mei:2013dk,Rofouei:2008tk,Collange:2009ci,DaQiRen:2010kc,Hong:2010ie,Jiao:2010cb,Nagasaka:2010cq,Kasichayanula:2012fp,Ge:2013ed}.
Our work differs in that we consider temperature, voltage and frequency
as independent variables over which to optimize performance-per-watt.
That is, we consider power efficiency $\eta_{\mathrm{pow}}=\eta_{\mathrm{pow}}(V_{\mathrm{dd},}f_{\mathrm{clock},}T)$.

While voltage and frequency have previously been explored in GPU DVFS
studies \citep{Mei:2013dk,Ge:2013ed,Nugteren:2014bo}, we explore
a larger parameter space. Apart from in Hong et. al. \citep{Hong:2010ie},
temperature effects on GPU power efficiency have been ignored. This
is detrimental to GPU power model accuracy and to achieving optimal
power efficiency, as discussed in Liao et. al. \citep{Liao:2005ft,Liao:2005ek}.
We show that the simplified linear model of Hong et. al. \citep{Hong:2010ie}
is not sufficient for predicting power usage on current generation
GPUs. To the authors knowledge, this is the first time the non-linear effect 
of temperature upon GPU efficiency has been studied in public literature.

The remainder
of this paper is organized as follows. In Section~\ref{sec:Materials-and-Methods},
we introduce the hardware and software used to find optimal power
efficiency on an NVIDIA K20 GPU. Our results are then presented in
Section~\ref{sec:Results}; this is followed by discussion (Section~\ref{sec:Discussion})
and conclusions (Section~\ref{sec:Conclusions}).

\section{Materials and Methods\label{sec:Materials-and-Methods}}


\subsection{Hardware overview}

The work presented here was conducted on ``GreenGPU'', a custom-built
computer system. GreenGPU consists of a Gigabyte GA-Z68MX motherboard
with an Intel i7-2600 CPU, 16~GiB of DDR3 RAM, and an NVIDIA Tesla
K20 GPU. The default heatsink of the K20 was replaced with an
EK-FCTK20 water block, and a Swiftech water cooling system (MCP655)
was installed. Water cooling was added to give access to a wider range of temperatures than possible using air cooling and to provide control of coolant flow.
The operating system used for
testing was 64-bit Linux Ubuntu 12.04 LTS, with NVIDIA GPU driver
version 319.37 installed. A Windows 7 partition was also installed
in order to run Windows-only GPU firmware modification tools.

\subsection{Clock and voltage management}

To control the clock frequency and voltage of the K20 GPU, we used
three tools: \texttt{\small{}nvidia-smi}%
\footnote{\href{https://developer.nvidia.com/nvidia-system-management-interface}{https://developer.nvidia.com/nvidia-system-management-interface}%
}, \texttt{\small{}GPU-Z}%
\footnote{\href{http://www.techpowerup.com/gpuz/}{http://www.techpowerup.com/gpuz/}%
} and\texttt{\small{} Kepler BIOS Tweaker}%
\footnote{\href{http://www.softpedia.com/get/System/Benchmarks/Kepler-BIOS-Tweaker.shtml}{http://www.softpedia.com/get/System/Benchmarks/Kepler-BIOS-Tweaker.shtml}%
}. The \texttt{\small{}nvidia-smi} utility, or NVIDIA System Management
Interface, is a command line utility that 
allows for the GPU core frequency to be altered; the allowed
values are dependent upon the GPU (Table~\ref{tab:nvidia-smi}).
The \texttt{\small{}nvidia-smi}
tool also allows for power draw and GPU die temperature to be read
from GPU sensors, giving an accurate way to measure temperature and
power, with differences between power and temperature reliable to
within $\pm$1~W and $\pm$$1^{\circ}$C. The reported power is the
full-board power consumption, which includes memory and voltage regulators. 

For finer grain control over core voltage and frequency, and so that
we could tune these as independent parameters, we used the \texttt{\small{}GPU-Z}
tool v0.7.7 and \texttt{\small{}Kepler BIOS Tweaker }tool v1.27. \texttt{\small{}GPU-Z}
is a utility that displays GPU specifications and operating parameters,
and allows for GPU firmware to be downloaded from the GPU. The \texttt{\small{}Kepler
BIOS Tweaker }tool allows for modification of the parameters within
GPU firmware, such as voltage and clock frequency. While benchmarking
was run on the Ubuntu partition of GreenGPU, these two programs were
run on the Windows 7 partition. Note that flashing firmware using
tools such as \texttt{\small{}Kepler BIOS Tweaker} will void warranty
and can potentially cause damage to the GPU. 

\begin{table}
\protect\caption{Supported K20 core and memory clock pairs \label{tab:nvidia-smi}}

\centering{}{\small{}}%
\begin{tabular}{cccc}
{\small{}GDDR5 Freq.} & {\small{}Core Freq.} & \multicolumn{2}{c}{{\small{}GPU Core Voltage }}\tabularnewline
{\small{} (MHz)} & {\small{} (MHz)} & {\small{}State ID} & {\small{}(mV)}\tabularnewline
\hline 
\hline 
{\small{}2600} & {\small{}758} & {\small{}V5} & {\small{}987.5-1112.5}\tabularnewline
 & {\small{}705} & {\small{}$\,$V4$^{*}$} & {\small{}950-1062.5}\tabularnewline
 & {\small{}666} & {\small{}V3} & {\small{}925-1050}\tabularnewline
 & {\small{}640} & {\small{}V2} & {\small{}912.5-1025}\tabularnewline
 & {\small{}614} & {\small{}V1} & {\small{}900-1000 }\tabularnewline
\hline 
{\small{}324} & {\small{}324} & {\small{}V0} & {\small{}875 - 875}\tabularnewline
\hline 
\multicolumn{4}{l}{{\footnotesize{}$^{*}$Default value}}\tabularnewline
\end{tabular}
\end{table}

\subsection{\texttt{xGPU} cross-correlation code}

For benchmarking and power efficiency testing, we used the \texttt{\small{}xGPU}
CUDA library\texttt{\small{}}%
\footnote{\href{https://github.com/GPU-correlators/xGPU}{https://github.com/GPU-correlators/xGPU}%
} \citep{Clark:2013fr}. \texttt{\small{}xGPU} computes the cross-correlation
of time-series data of $N$ inputs and is used for interferometric
synthesis imaging in radio astronomy, see for example \citep{Kocz:2014jr}.
It is virtually identical to the BLAS routine CHERK --- Complex Hermitian
Rank K update --- where the $T\times N$ matrix, corresponding to
time series data ($T$ dimension) from $N$ antennas is multiplied
by its complex conjugate, producing an $N\times N$ Hermitian matrix.
The problem is compute-bound because the compute complexity scales
as $N^{2}T$, whereas the memory traffic scales as $N(T+N)$. Results presented
here used values  $N=8192$ and $T=1000$.
The \texttt{\small{}xGPU} code differs from cuBLAS
CHERK as it contains domain-specific tweaks: it is designed to process
8-bit integer input (processed as 32-bit floating point), only stores the lower triangle of the correlation
matrix, and uses smaller tiles to improve performance for small-$N$.
\texttt{\small{}xGPU} also has an additional parameter corresponding
to the number of frequency channels to process; the problem is trivially
parallelizable over frequency channels, so this can be thought of
as a batching parameter. 

Here, we use the \texttt{\small{}xGPU} application because that is our domain of interest; 
however, we note that given it is compute-bound, it is well suited to our investigation:
when running this
algorithm, most of the power is consumed by the floating point units,
and this increases the validity of the simple model in Section 1.2.

\texttt{\small{}xGPU} has two different modes that were of particular
use for this work. The first mode is a benchmark, which computes various
performance metrics achieved, such as FLOPS, for a given set of compile-time
parameters. The output of the GPU code is also compared against CPU
code for validation. The second mode is a power diagnostic loop, in
which \texttt{\small{}xGPU} is fed dummy data and run in an infinite
loop, so as to keep the GPU running continuously.

For the compile-time parameters used, 
the single-precision computational performance $p$ in FLOPS for \texttt{\small{}xGPU}
was found to follow $p=2.89 f_{MHz}$, for
clock frequencies between 614-1070~MHz, with a maximum performance of
3094~GFLOPS at 1070~MHz. Note that temperature and
voltage do not affect achieved FLOPS. 


\subsection{Performance profiling method}

The main parameters used for testing power efficiency in this work
were GPU die temperature, GPU core voltage and clock frequency. We
used \texttt{\small{}Kepler BIOS Tweaker }and \texttt{\small{}nvidia-smi}
to modify the GPU core voltage and clock frequency, then we used \texttt{\small{}xGPU}
to benchmark performance. Attempts to vary the memory clock frequency
resulted in the GPU being inoperable, so no memory clock adjustments
were conducted. 

Thermal control of the GPU die was achieved by running \texttt{\small{}xGPU}
in a power loop, while controlling the flow of water through the water
cooling system. In order to continuously monitor the temperature and
power draw, we used a Python script to parse the output of \texttt{\small{}nvidia-smi
}and to log timestamped power usage and temperature data to file every
second. By running this script in tandem with \texttt{\small{}xGPU},
we tested the performance of the K20 GPU over a variety of core frequency
and voltage combinations.

\section{Results\label{sec:Results}}




\subsection{Overclocking at constant temperature}

\begin{figure*}
\begin{centering}
\subfloat[\label{fig:pow-vs-freq}Measured GPU power usage.]{\begin{centering}
\includegraphics[width=0.80\columnwidth]{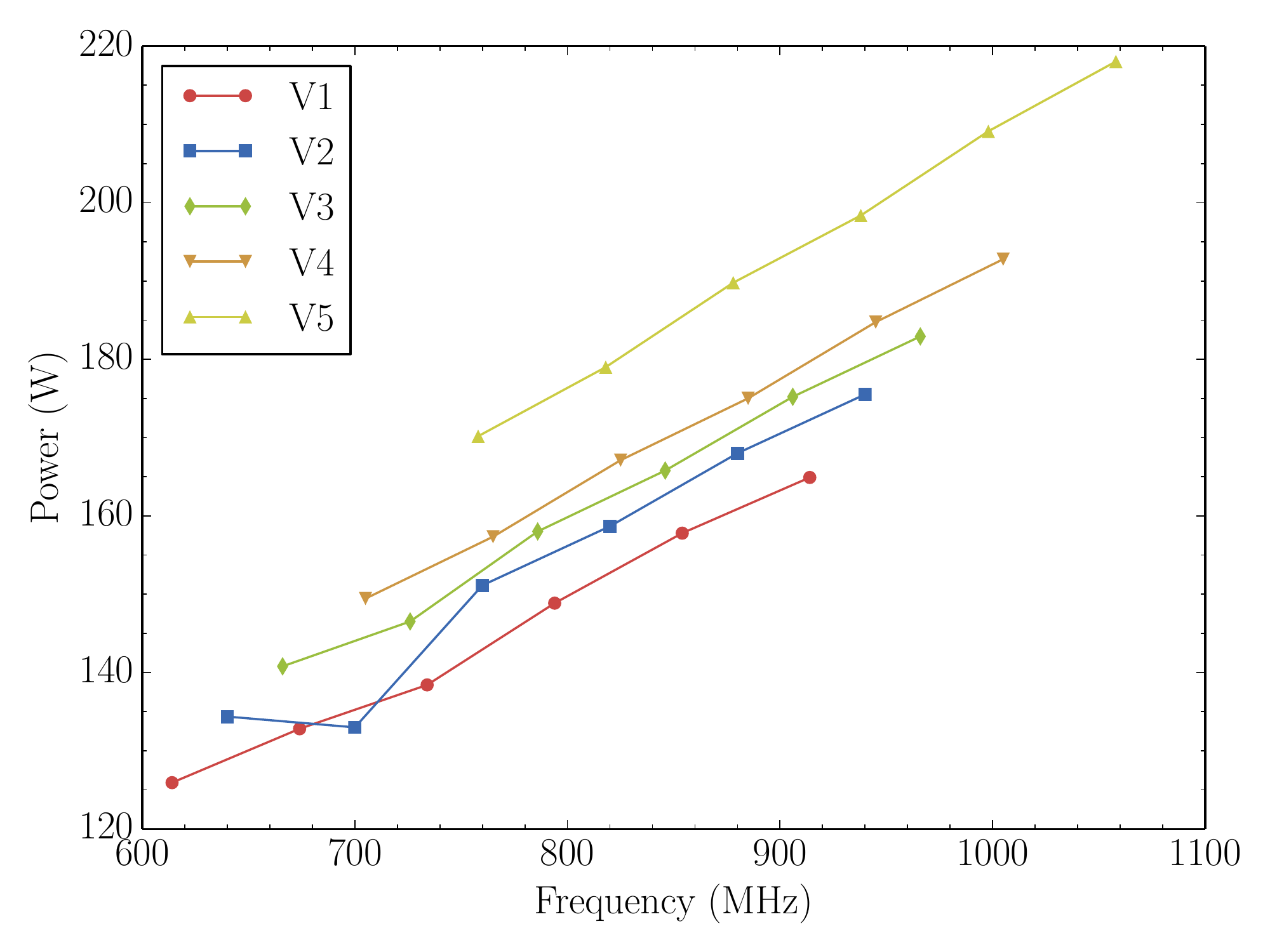}
\par\end{centering}
}\subfloat[\label{fig:ppw-vs-freq}Measured GPU power efficiency.]{\begin{centering}
\includegraphics[width=0.80\columnwidth]{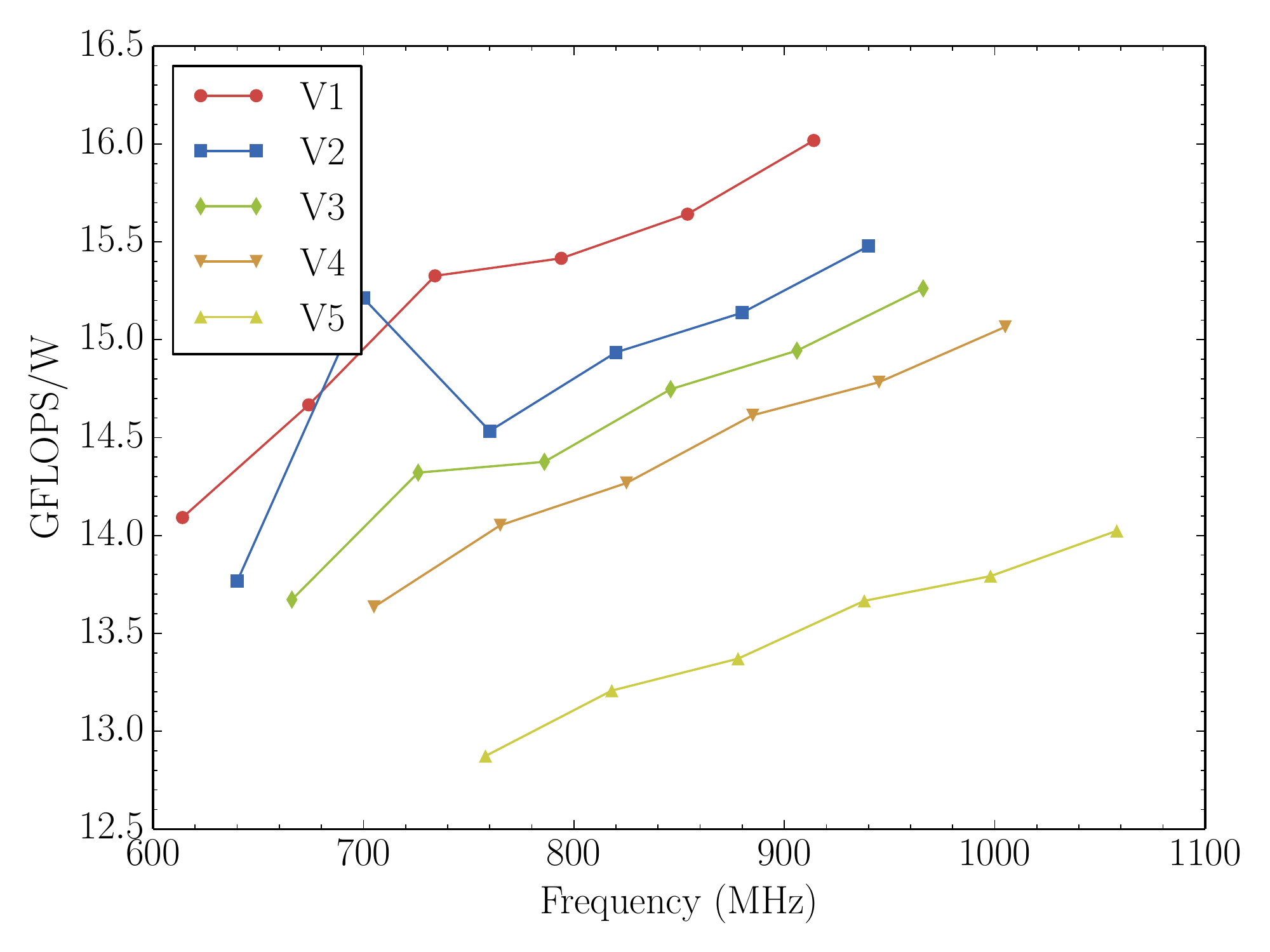}
\par\end{centering}
}
\protect\caption{GPU power usage and efficiency for \texttt{xGPU} code running on a
K20 GPU, for default voltages (V1-V5) with frequency offsets of 0-300
MHz over default $f_{\mathrm{clock}}$ settings (see Table \ref{tab:nvidia-smi}). }
\end{centering}
\end{figure*}

After profiling the computational performance of \texttt{\small{}xGPU},
we compared power usage of the GPU at different ($f_{clock}$, $V_{dd}$)
combinations. As shown in Table~\ref{tab:nvidia-smi}, the K20 has
preset frequency-voltage combinations that can be selected with \texttt{\small{}nvidia-sm}i.
We applied frequency offsets of 0-300 MHz to these default values,
in 60~MHz increments, and then measured the resulting power usage
for the \texttt{\small{}xGPU }code (Fig.~\ref{fig:pow-vs-freq}),
and the corresponding power efficiency (Fig.~\ref{fig:ppw-vs-freq}).
Voltage states are labelled V1-V5, with increasing voltage; for these
data, the firmware voltage table was not modified. To account for
temperature effects, we held GPU die temperature at 34$\pm$2$^{\circ}$C. 

The default voltage state (V4) with default  $f_{\mathrm{clock}}$ of
705~MHz yields a power efficiency of 13.6~GFLOPS/W. We find a peak
power efficiency of 16.0 GFLOPS/W when using
the lowest voltage state with a $f_{\mathrm{clock}}$ of 914~MHz, an
increase of 18\%. The worst power efficiency was achieved when using
the highest voltage level with its default $f_{\mathrm{clock}}$ of 758~MHz.
The dip in power usage at $f_{\mathrm{clock}}$= 705~MHz when in
the V2 state is likely due to the GPU selecting a low core voltage
within the allowed range (see Table~\ref{tab:nvidia-smi}).

\subsection{Temperature dependence of power efficiency\label{sub:Overclock-const-v}}

\begin{figure}
\begin{centering}

\includegraphics[width=0.8\columnwidth]{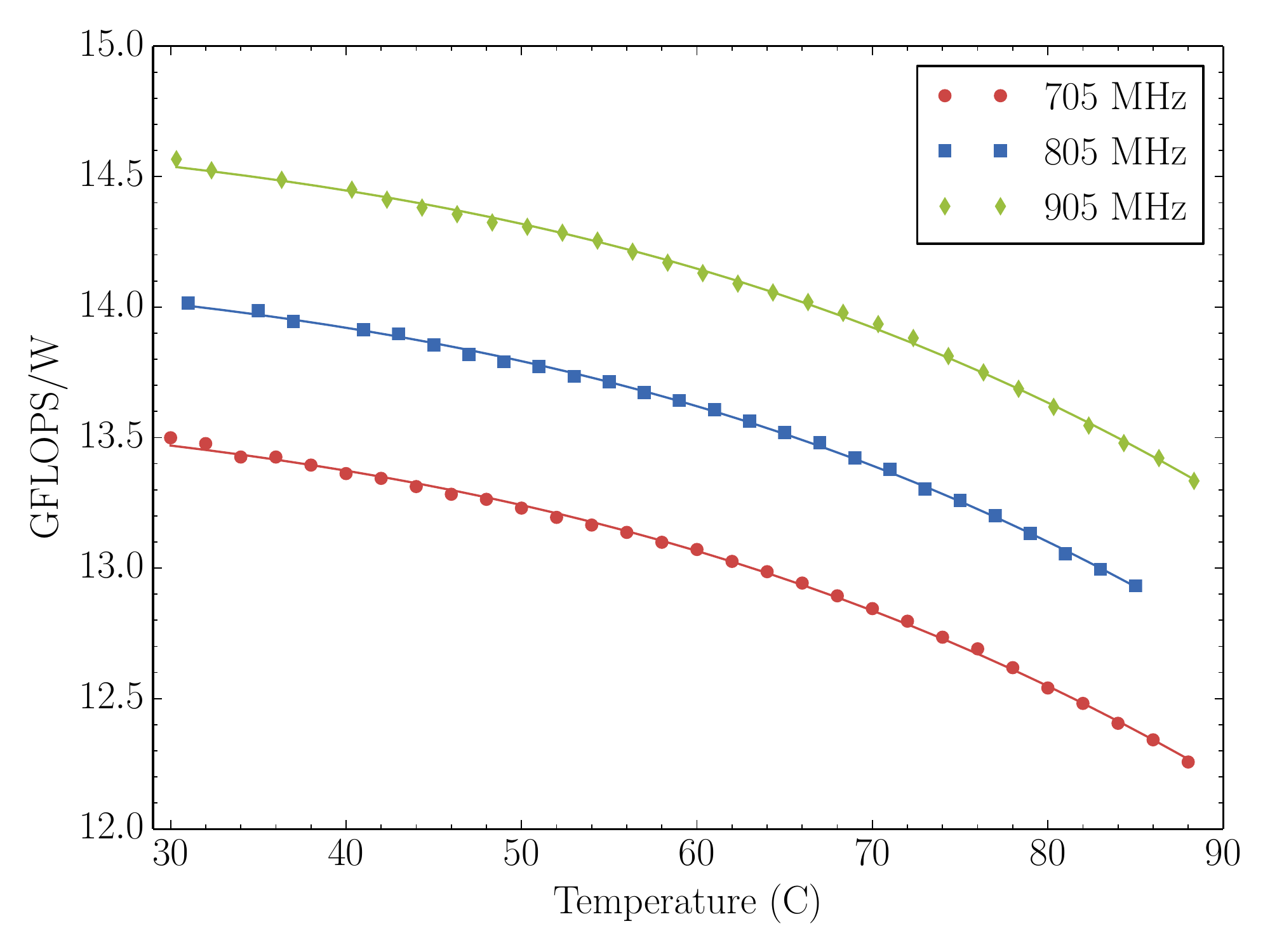}


\protect\caption{ \texttt{xGPU} efficiency on a
K20 GPU, with default core voltage (V4). Power usage shows a strong
non-linear temperature dependence; this decreases performance per
watt as temperature increases. Here,
$T=30^{\circ}$C. \label{fig:const-v-ppw} }
\end{centering}
\end{figure}

Eq. \ref{eq:sub-thr} predicts that subthreshold leakage current
is proportional to $T^{2}e^{-b/T}$. To investigate this, we compared
power efficiency of the GPU at various die temperatures,where
we have averaged multiple data into bins of $\pm1^{\circ}$C (Fig.~\ref{fig:const-v-ppw}),
the clock frequency was set to 705, 805 and 905~MHz, with the default
core voltage state (950-1062.5~mV). At all temperatures, power usage
changes by a fixed $\sim0.14$~W/MHz. As clock frequency does not
affect static power $P_{static}$, the offset between lines corresponds
to the dynamic power $P_{\mathrm{dynamic}}$ component of the total
power usage.

We also see a non-linear increase of power consumption as a function
of temperature; the simple linear model as presented in \citep{Hong:2010ie}
is not sufficient. If we take into account $P_{\mathrm{dynamic}}$,
we can fit a model , $P_{\mathrm{static}}=aT^{2}e^{-b/T}+c$ to all
three runs (solid lines). For temperature in Kelvin, a least-square
fit yields $a=1.00\pm0.23$, $b=3209.7\pm83.7$, $c=(148.9\pm0.2,\,162.7\pm0.2,\,176.9\pm0.2)$
for 705, 805 and 905 MHz, respectively. 

Power efficiency is improved as clock frequency is increased. There is an 18\% difference
in power efficiency between worst (705~MHz at 90$^{\circ}$C) and best (905~MHz at 30$^{\circ}$C) cases.
At constant $T$=30$^{\circ}$C, the performance at 905~MHz is 14.6 GFLOPS/W,
as opposed to 13.5 GFLOPS/W at 705~MHz; an 8.1\% increase.

\subsection{Constant frequency, modified voltage}

\begin{figure}
\begin{centering}

\includegraphics[width=0.8\columnwidth]{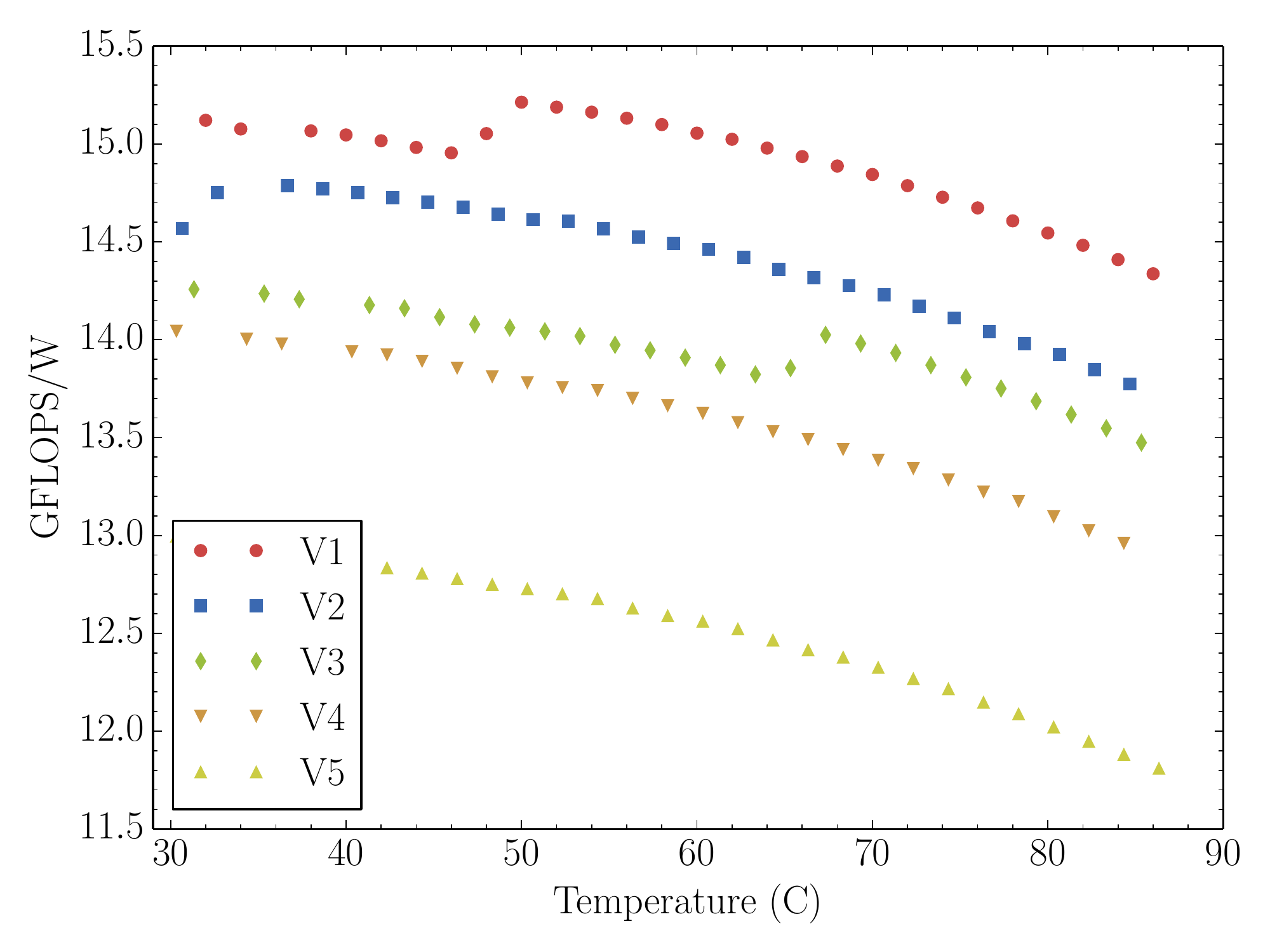}


\protect\caption{ \texttt{xGPU} power usage and efficiency on a
K20 GPU, with a core frequency of 800 MHz and varying voltage
levels (see Table \ref{tab:nvidia-smi})\label{fig:const-f-ppw}. Here,
$T=30^{\circ}$C.}
\end{centering}
\end{figure}

The default voltage states of the K20 are not fixed voltages, but
rather a range (Table~\ref{tab:nvidia-smi}). To investigate the
effect of voltage on power efficiency, we reprogrammed the K20's firmware
so that the GPU core voltages V1-V5 were fixed to the lower bound of the default voltage ranges (Table~\ref{tab:nvidia-smi}).
Power efficiency as a function of temperature for the modified voltage
levels V1-V5 is shown in Fig.~\ref{fig:const-f-ppw}, for a constant
clock frequency of $f_{\mathrm{clock}}=$800~MHz. 
As voltage is increased, power efficiency decreases. 
The highest efficiency of 14.7 GFL-OPS/W was achieved using the V1 state, 
while the V5 state yielded 12.6 GFLOPS/W, the lowest for these tests. 
This corresponds to a 16.7\% difference in power
efficiency between best and worst cases.

Apparent in Fig.~\ref{fig:const-f-ppw} are unexpected discontinuous jumps in the 
reported power usage. These
drops are repeatable and occur at different temperatures for different
voltage states. We are uncertain as to the cause; however, \texttt{nvidia-smi}
does not report clock throttling and no decrease in performance is seen. 
The altered voltage table (as written in the GPU's firmware) did not allow for different voltage states, 
and the K20 GPU does not employ temperature-dependent voltage scaling. 
We conclude that this is due to an unknown off-chip (i.e. off-processor)
effect. A possible explanation is that this is due to current-dependent
efficiencies in power delivery of the regulators that supply the GPU
die with power.

\subsection{Tuning voltage and frequency\label{sub:best-ppw}}

The best performance-per-watt is achieved when undervolting and overclocking
the GPU, as predicted in Section~\ref{sub:Maximizing-power-efficiency}
(Table~\ref{tab:pow-eff}). At 900~mV, \texttt{xGPU} code execution
fails and the GPU froze when attempting to run the code at 1005~MHz.
At 955~MHz, the code ran successfully but the output failed verification
when GPU temperature was above 70$^{\circ}$C; that is, it did not
match the output of reference CPU code. No numerical errors were found for temperatures
below 70$^{\circ}$C. At 875~mV, we achieved a maximum clock frequency
of 905~MHz, but again found that GPU output did not pass verification
for temperatures above 70$^{\circ}$C. 

For $(V,\, f_{\mathrm{clock}},T)$ = (875~mV, 905~MHz, 30$^{\circ}$C),
we achieved 18.3~GFLOPS/W for the \texttt{xGPU} code. For comparison,
the K20 default of $(V,\, f_{\mathrm{clock}},T)$ = (950-1062.5 mV,
705~MHz, 30$^{\circ}$C) yields 13.5~GFLOPS/W for the same code,
degrading to 12.4~GFLOPS/W at 90$^{\circ}$C. This means that by
controlling GPU temperature, voltage, and clock frequency, we are
able to increase performance-per-watt by 37-48\% over default settings.\captionsetup{position=top}

\section{Discussion\label{sec:Discussion}}

Our results show that temperature has a nontrivial impact on GPU power
efficiency. This is primarily due to leakage current, which scales
in proportion to $T^{2}e^{-b/T}$. Optimal power efficiency is achieved
with lowest possible GPU supply voltage with highest possible  clock frequency
at low temperature. We find efficiency can be increased
by as much as 48\% on an NVIDIA K20 through this technique. 

We have demonstrated a $\sim$30~W decrease on a GPU power consumption
of 154~W by changing GPU core voltage state, a 20\% reduction (Table~2).
Coupled with an increase in GPU clock frequency, 
performance increased from 2064 GFLOPS to 2636 GFLOPS --- a 28\%
increase --- while simultaneously  power usage dropped by 10~W. 
For large installations, even a small change in power efficiency can
have significant cost benefits. 

%

We have presented results from a single GPU. In actuality, the same
chips from within the same process will have a distribution of values
(dynamic power, leakage power, etc.), so a degree of conservatism
is required in setting device parameters for mass production. Allowing
clock frequency and core voltage to be set at run-time by the user,
or adjusted automatically using dynamic voltage and frequency scaling
techniques (DVFS), may provide a mechanism with which to boost power
efficiency over conservative defaults.

\begin{table}
\begin{centering}
\protect\caption{Power efficiency (at 50$^{\circ}$C) and benchmarks for \texttt{xGPU}
code as a function of voltage and clock frequency.\label{tab:pow-eff}.}
{\centering{}{\footnotesize{}}%
\begin{tabular}{ccccc}
{\footnotesize{} Freq.} & {\footnotesize{}Voltage} & {\footnotesize{}Power} & {\footnotesize{}Benchmark} & {\footnotesize{}Power eff.}\tabularnewline
{\scriptsize{}(MHz)} & {\scriptsize{}(mV)} & {\scriptsize{}(W)} & {\scriptsize{}(GFLOPS)} & {\scriptsize{}(GFLO-}\tabularnewline
{\scriptsize{}} & {\scriptsize{}} & {\scriptsize{}} & {\scriptsize{}} & {\scriptsize{}PS/W)}\tabularnewline
\hline 
\hline 
{\footnotesize{}1070} & {\footnotesize{}987.5} & {\footnotesize{}222.2} & {\footnotesize{}3094} & {\footnotesize{}13.9}\tabularnewline
{\footnotesize{}1017} & {\footnotesize{}950.0} & {\footnotesize{}197.1} & {\footnotesize{}2940} & {\footnotesize{}14.9}\tabularnewline
{\footnotesize{}978} & {\footnotesize{}925.0} & {\footnotesize{}186.9} & {\footnotesize{}2826} & {\footnotesize{}15.1}\tabularnewline
{\footnotesize{}952} & {\footnotesize{}912.5} & {\footnotesize{}175.6} & {\footnotesize{}2750} & {\footnotesize{}15.7}\tabularnewline
{\footnotesize{}926} & {\footnotesize{}987.5} & {\footnotesize{}199.0} & {\footnotesize{}2674} & {\footnotesize{}13.4}\tabularnewline
{\footnotesize{}926} & {\footnotesize{}950.0} & {\footnotesize{}183.1} & {\footnotesize{}2674} & {\footnotesize{}14.6}\tabularnewline
{\footnotesize{}926} & {\footnotesize{}925.0} & {\footnotesize{}179.4} & {\footnotesize{}2674} & {\footnotesize{}14.9}\tabularnewline
{\footnotesize{}926} & {\footnotesize{}912.5} & {\footnotesize{}172.0} & {\footnotesize{}2674} & {\footnotesize{}15.5}\tabularnewline
{\footnotesize{}926} & {\footnotesize{}900.0} & {\footnotesize{}168.5} & {\footnotesize{}2674} & {\footnotesize{}15.9}\tabularnewline
{\footnotesize{}905} & {\footnotesize{}default} & {\footnotesize{}181.6} & {\footnotesize{}2636} & {\footnotesize{}14.5}\tabularnewline
{\footnotesize{}905} & {\footnotesize{}875.0} & {\footnotesize{}144.4} & {\footnotesize{}2636} & {\footnotesize{}18.1$^{a}$}\tabularnewline
{\footnotesize{}805} & {\footnotesize{}default} & {\footnotesize{}167.5} & {\footnotesize{}2330} & {\footnotesize{}13.9}\tabularnewline
{\footnotesize{}805} & {\footnotesize{}875.0} & {\footnotesize{}132.6} & {\footnotesize{}2330} & {\footnotesize{}17.5}\tabularnewline
{\footnotesize{}705} & {\footnotesize{}default} & {\footnotesize{}153.9} & {\footnotesize{}2064} & {\footnotesize{}13.4}\tabularnewline
{\footnotesize{}705} & {\footnotesize{}875.0} & {\footnotesize{}122.1} & {\footnotesize{}2064} & {\footnotesize{}16.9}\tabularnewline

\hline 
\multicolumn{5}{c}{{\scriptsize{}$^{a}$output from the GPU does not pass validation
for $T>$70$^{\circ}$C.}}\tabularnewline
\end{tabular}}
\par\end{centering}

\end{table}
\captionsetup{position=bottom}

\subsection{Application and temperature-aware DVFS}

When a manufacturer chooses clock frequencies for a GPU, 
it is typical to choose
clock frequencies that can support a wide range of workloads.
For example, codes such as \texttt{DGEMM} (double-precision general
dense matrix multiply) consume more power than the single-precision
\texttt{xGPU} code, but must still run within the TDP at default clock
frequency. It follows that there will always be a significant boost
in clock frequencies possible for applications that do not run close
to the TDP limit. One could imagine a control system that automatically
adjusts clock frequency and voltage, depending upon application and
desired performance optimization (e.g. FLOPS/W or FLOPS). This would
be a form of DVFS. Such an application-dependent frequency and voltage
scaling system\emph{ (A-DVFS}) could offer a way to automatically
boost power efficiency and performance of codes.
This approach could also accommodate
applications that require perfect load balancing or reduced system
jitter by setting clock frequencies uniformly across all devices used
by the application.

Indeed, the NVIDIA GPU Boost feature%
\footnote{\href{http://www.nvidia.com/content/PDF/kepler/nvidia-gpu-boost-tesla-k40-06767-001-v02.pdf}{http://www.nvidia.com/content/PDF/kepler/nvidia-gpu-boost-tesla-k40-06767-001-v02.pdf}%
}, launched with the K40 series GPU, allows users to select from two
preset higher clocks through \texttt{\small{}nvidia-smi}, boosting
performance for codes that run below TDP. GPU Boost is implemented
differently on the GeForce-class gaming cards: core frequency is scaled
to maintain card power consumption close to TDP. Adding similar dynamic
frequency scaling functionality to server-class GPUs may increase
both power efficiency and performance for codes with low power consumption.

Temperature and TDP limit the range of clock frequencies and supply
voltage combinations. The default settings for GPUs are chosen specifically
to ensure that neither temperature or TDP tolerances are exceeded
for any application. In contrast, best power efficiency occurs when
voltages are lowered and clock frequency raised in accordance with
operating temperature.  A hypothetical temperature-aware voltage and frequency scaling system
\emph{(T-DVFS}) could raise and lower core voltages automatically,
based on the GPU die temperature. If cooling systems maintained lower
temperatures, the T-DVFS system would accordingly lower voltage, increasing
power efficiency.

\subsection{Cooling Systems}

Our water-cooling system allowed us to operate the GPU at lower die temperatures
under load than that possible with the stock fan. Overall power efficiency of a GPU-based 
HPC system
depends also on the power consumed by cooling subsystems.
Our water-based system used less power than the chassis' stock fans, so in our simple case
overall power efficiency increased. In larger installations,
Januszewskia et al. report that water-based cooling systems can reduce the total power consumed
by a server room by more than 15\% \citep{Januszewskia:vu}. 
Warm water-based cooling techniques show great promise; 
an IBM Aquasar system demonstrated
an exergetic efficiency increase of 34\% through use of warm water
(60$^{\circ}$C) cooling \citep{Zimmermann:2012fv}. However,  power
consumption of electronics increased by $7\pm1\%$ as the coolant
temperature increases from 30$^{\circ}$C to 60$^{\circ}$C. 



If we modify Eq. \ref{eq:pow-eff} to include
the power required for cooling $P_{\mathrm{cool}}$ and other infrastructure sources, we instead
wish to optimize
\begin{equation}
\eta_{\mathrm{pow}}(V,f,T)=\frac{N_{\mathrm{OPS}}(V,f,T)}{P_{\mathrm{sys}}(V,f,T)+P_{\mathrm{cool}}(T)+...},\label{eq:optimize-eff-temp}
\end{equation}
where the denominator is the sum of the power over the entire system.
Here, we have explicitly written $P_{\mathrm{sys}}$ and $P_{\mathrm{cool}}$
as functions of temperature. Using Eq.~\ref{eq:optimize-eff-temp}
as a basis for finding optimal power efficiency for a given code differs
from past techniques as it considers the system as a whole, with regards
to the fundamental physics that governs power usage of the underlying
microarchitecture.

A novel aspect of Eq.~\ref{eq:optimize-eff-temp} is that it
predicts that lowering temperature may lead to increased power efficiency,
which appears somewhat in conflict to previous findings that report
lower data center energy consumption at higher temperatures. There are two main reasons
this discrepancy arises. Firstly, general-purpose data centers focus
on optimizing power usage effectiveness (PUE),
as opposed to performance-per-watt, which is of more interest to HPC
systems. 
PUE is defined as the ratio of  total facility energy (data center's total energy usage)
to IT equipment energy (sum of all computing, storage and network
equipment energy usage). 
Unlike performance-per-watt, PUE does not
directly consider the computational performance of a system. Secondly,
previous comparisons between cooling methods do not account for temperature-dependent
optimization of supply voltage and clock frequency.

\section{Conclusions\label{sec:Conclusions}}

One of the main challenges facing exascale HPC is dramatically reducing
the power usage of large HPC systems. We have shown that temperature-aware
optimization of core clock frequency and supply voltage can increase
performance of a GPU code by up to 48\% on an NVIDIA Tesla K20,
achieved by increasing the GPU clock frequency and decreasing
supply voltage while maintaining a die temperature of 30$^{\circ}$C.

It is taken for granted that code must be optimized for different
architectures in order to fairly compare compute performance. In contrast,
when optimizing power efficiency for HPC systems, the effect of temperature
upon optimal GPU core frequency and voltage is generally not considered.
Temperature-aware and application-dependent frequency and
voltage scaling (T-DVFS and A-DVFS) may provide a mechanism with which
to increase the power efficiency of GPUs for HPC, by automatically tuning
 frequency and voltage with consideration of both application
code and thermal environment.


\bibliographystyle{spmpsci}      
\bibliography{greengpu}

\end{document}